\documentclass[twocolumn,showpacs,superscriptaddress,prl,aps,floatfix]{revtex4}

\usepackage{graphicx}
\usepackage{rotating}
\usepackage{amsmath}
\usepackage[usenames,dvipsnames]{color}

\begin{document}

\title{Polaronic hole-trapping in doped $\rm BaBiO_3$}

\author{C. Franchini}

\affiliation{
Faculty of Physics, Universit\"at Wien and Center for Computational Materials Science,  A-1090 Wien, Austria}

\author{G. Kresse} 
\affiliation{
Faculty of Physics, Universit\"at Wien and Center for Computational Materials Science,  A-1090 Wien, Austria}

\author{R. Podloucky}      
\affiliation{
Institute for Physical Chemistry and Center for Computational Materials 
Science, Vienna University, Sensengasse 8, A-1090 Vienna, Austria}

\date{\today}
\pacs{71.30.+h,71.38.-k,71.45.Lr,78.20.-e}

\begin{abstract}
{ 
The present {\em ab initio} study shows that in BaBiO$_3$, Bi$^{3+}$ sites can trap two holes from the valence band
to form Bi$^{5+}$ cations. The trapping is accompanied by large local lattice distortions, therefore
the composite particle consisting of the electronic-hole and the local lattice phonon field forms a polaron.
Our study clearly shows that even $sp$ elements can trap carriers at lattice sites,
if local lattice relaxations are sufficiently large to screen the localised hole.
The derived model describes all relevant experimental results, and settles the issue of
why hole doped BaBiO$_3$ remains semiconducting upon moderate hole doping.
}
\end{abstract}

\maketitle

\begin{figure}
\includegraphics[clip,width=0.49\textwidth]{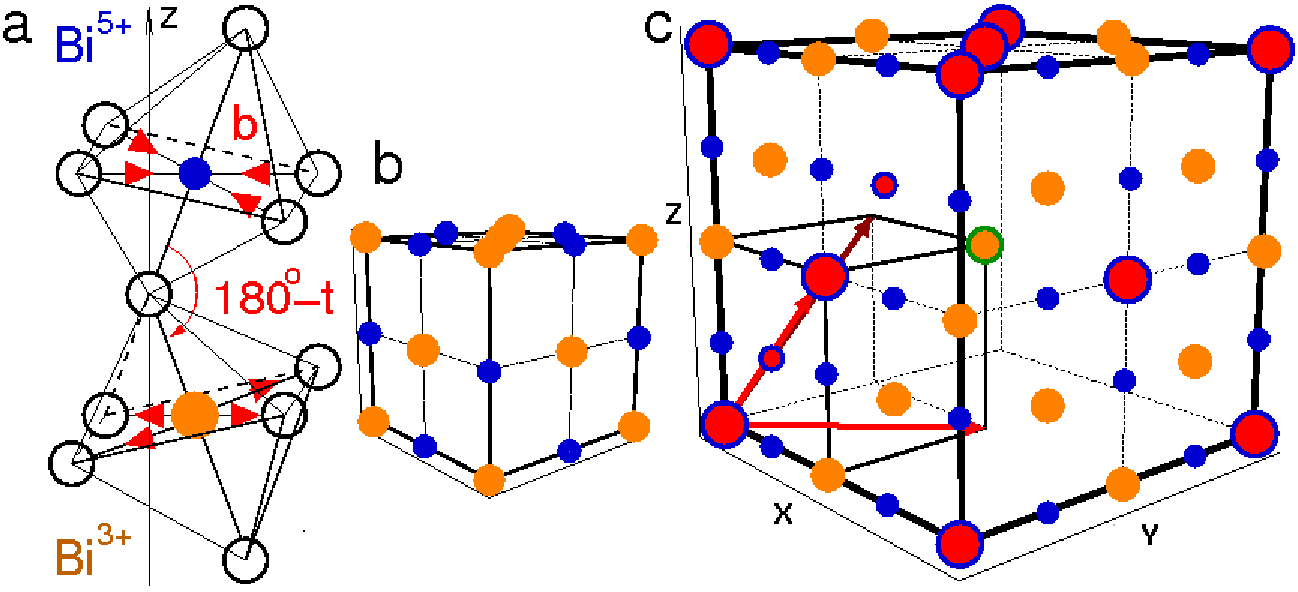}
\caption{
(a) Monoclinically distorted $\rm BiO_6$ octahedra emphasising the tilting ($t$) and
breathing ($b$) structural instabilities around inequivalent $\rm Bi^{3+}$ (orange) and $\rm Bi^{5+}$ (blue) atoms.
(b) Sketch of the simple cubic (sc) small superlattice comprising 8 $\rm BiO_6$ octahedra. 
(c) Representation of the large face centred cubic (fcc) supercell (16 $\rm BiO_6$ octahedra) adopted 
to model the polaronic states for
$x=0.125$ and $0.25$,
with red arrows indicating the Bravais lattice vectors. 
Red atoms indicate $\rm Bi^{3+}$ ions converted into $\rm Bi^{5+}$ ions upon hole-doping
for $x=0.125$ (large balls) and $x=0.25$ (large and small balls).
}
\label{fig:1}
\end{figure}

The nature of insulator-metal(superconductor) transitions in oxides has intrigued 
scientists for several decades\cite{tokura, mott, imada}. Such transitions can be realised  
by modulating the electrical carrier density using electric field effects, chemical
doping or pressure\cite{ahn,ueno}. Despite the intense experimental and theoretical efforts,
an atomistic interpretation of the experimental data is often exceedingly difficult, since
the available first principles methods are too approximate.
$\rm BaBiO_3$ is 
a prime example for such a phenomenon because it is a 
charge-ordered insulator, in which bismuth atoms appear in two different oxidation states (3+ and 5+).
It undergoes an insulator-superconductor transition upon hole doping\cite{cava}. 
Here we demonstrate that up to 0.25 holes per formula unit, the charge-ordered insulating state
prevails, but an insulator to metal transition is predicted beyond a hole concentration of
about 0.3 holes per formula unit. We show that the transition is controlled by the coupling
between the additional holes trapping on original Bi$^{3+}$ sites and the oxygen polarisation fields that surround
the BiO$_6$ octahedra. The composite particle, composed of a hole plus its accompanying 
polarisation field (distortion) is known as a polaron\cite{ziman}

The primitive cell of $\rm BaBiO_3$ can be described as Ba$_2^{2+}$Bi$^{3+}$Bi$^{5+}$O$^{2-}_6$, 
where $\rm Bi^{5+}$ and $\rm Bi^{3+}$ cations occur in equal parts\cite{cox,harrison}.
Indeed, the two different valencies of bismuth are manifested in the measured structural properties,
from which two different bismuth-oxygen bond lengths are derived: the shorter bond length is attributed to Bi$^{5+}$-O bonds,  
the larger one to Bi$^{3+}$-O bonds\cite{cox}.
At low temperatures, the two Bi species are alternatingly ordered 
in a distorted cubic (monoclinic) structure, in which Bi$^{5+}$ is surrounded by Bi$^{3+}$ neighbours (and vice versa),
as sketched in Fig. \ref{fig:1}(a) and (b).
Upon K doping, Ba$_{1-x}$K$_x$BiO$_3$ undergoes a phase transition to an orthorhombic, still perovskite like 
semiconducting phase, at a composition of $x \approx 0.12$\cite{pei}. 
Finally for larger K concentrations $0.37<x<0.53$, the compound becomes superconducting\cite{cava} 
with no measurable distortions.                               
{\em Ab initio} modelling for such a case
is problematic, since standard semi-local exchange-correlation functionals, such as the local density approximation
or the generalised gradient approximation, predict a too small  charge disproportionation
describing $\rm BaBiO_3$ as a metal\cite{mattheiss,kunc,rabe}, in disagreement with the experimental observations\cite{tajima}.
Since already the parent material is described incorrectly by semi-local functionals, 
density functional theory (DFT) calculations for hole doped $\rm BaBiO_3$ are unreliable, and until today an atomistic
understanding of why $\rm BaBiO_3$ remains semiconducting upon replacing the divalent 
Ba by monovalent K is missing. 

The drawbacks of semi-local density functionals are mainly related to sizeable self-interactions\cite{selfenergy, liech}. 
A theory that is largely self-interaction free is therefore a prerequisite to
model $\rm BaBiO_3$, and for this, hybrid functionals mixing a  fraction
of non-local Hartree-Fock exchange (typically 25 \%) to the otherwise semi-local exchange
are a promising choice\cite{becke}. In the present work, all calculations are based on the Heyd-Scuseria-Ernzerhof 
(HSE) hybrid functional\cite{heyd} and were carried out using the Vienna {\em ab-initio} simulation package (VASP)\cite{vasp,hse6}. 
This functional can be applied to metallic systems, since the long range tail of the Coulomb kernel is screened.

\begin{figure}
\centering
\includegraphics[clip=true,width=0.95\linewidth]{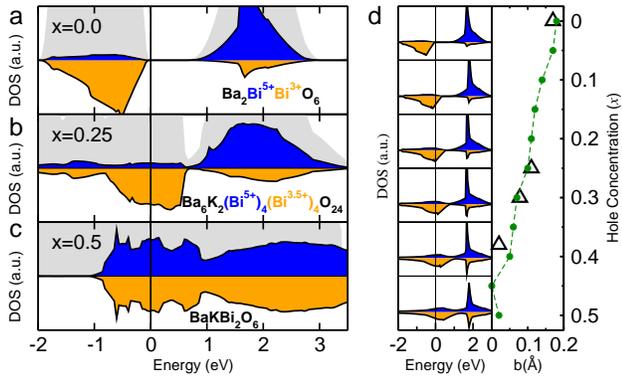}
\caption{(Colour online)
(a-c) Evolution of the density of states (DOS) on $\rm Bi^{3+}$ (orange) and $\rm Bi^{5+}$ (blue) atoms with K hole doping $x$
for the non-polaronic solution (superlattice with 8 unit cells).
The gray shadows indicate the total DOS.
(d) Evolution of the DOS of hole-doped $\rm BaBiO_3$ simulated by removing a fraction of an electron ($x$) leading to a
reduction of the gap between the $\rm Bi^{3+}$ and $\rm Bi^{5+}$ sub-bands,
and the simultaneous reduction of the Bi-O bond length difference $b$ between $\rm Bi^{3+}$  and $\rm Bi^{5+}$ sites.
Symbols $\bigtriangleup$ represent the estimated values of the breathing distortion $b$ 
computed from the Rice and Wang gap equation\cite{karlow}.
}
\label{fig:2}
\end{figure}

For pure BaBiO$_3$, HSE correctly predicts a semiconducting state 
with an indirect band gap of 0.65 eV [see Fig. \ref{fig:2}(a)], exhibiting Bi$^{3+}$ and  Bi$^{5+}$ sites.
For these two sites, the Bi-O bond lengths differ by $b$=0.18~\AA, which is in excellent
agreement with experiment($b$=0.17~\AA)\cite{cox}. 
The calculated tilting instability $t$=11.9$^{\circ}$
correlates also very well with the measured value of 11.2$^{\circ}$\cite{cox}.
The density of states shown in Fig. \ref{fig:2}(a) clearly shows that the valence band is dominated 
by orbitals located predominantly on Bi$^{3+}$ atoms and the surrounding oxygen atoms, whereas the 
conduction band is dominated by orbitals located close to the  Bi$^{5+}$ atoms.
It should be noted that we keep the simple formal notation of $3+$ and $5+$ for the Bi-ions, although
the actual charge transfer is reduced by screening, hybridisation and back-donation 
as a result of the selfconsistent treatment. 
We will first concentrate on the hole doped system without polaronic distortion,
and then discuss how polaronic distortions modify the electronic and lattice structure.

\begin{figure}
\includegraphics[clip,width=0.45\textwidth]{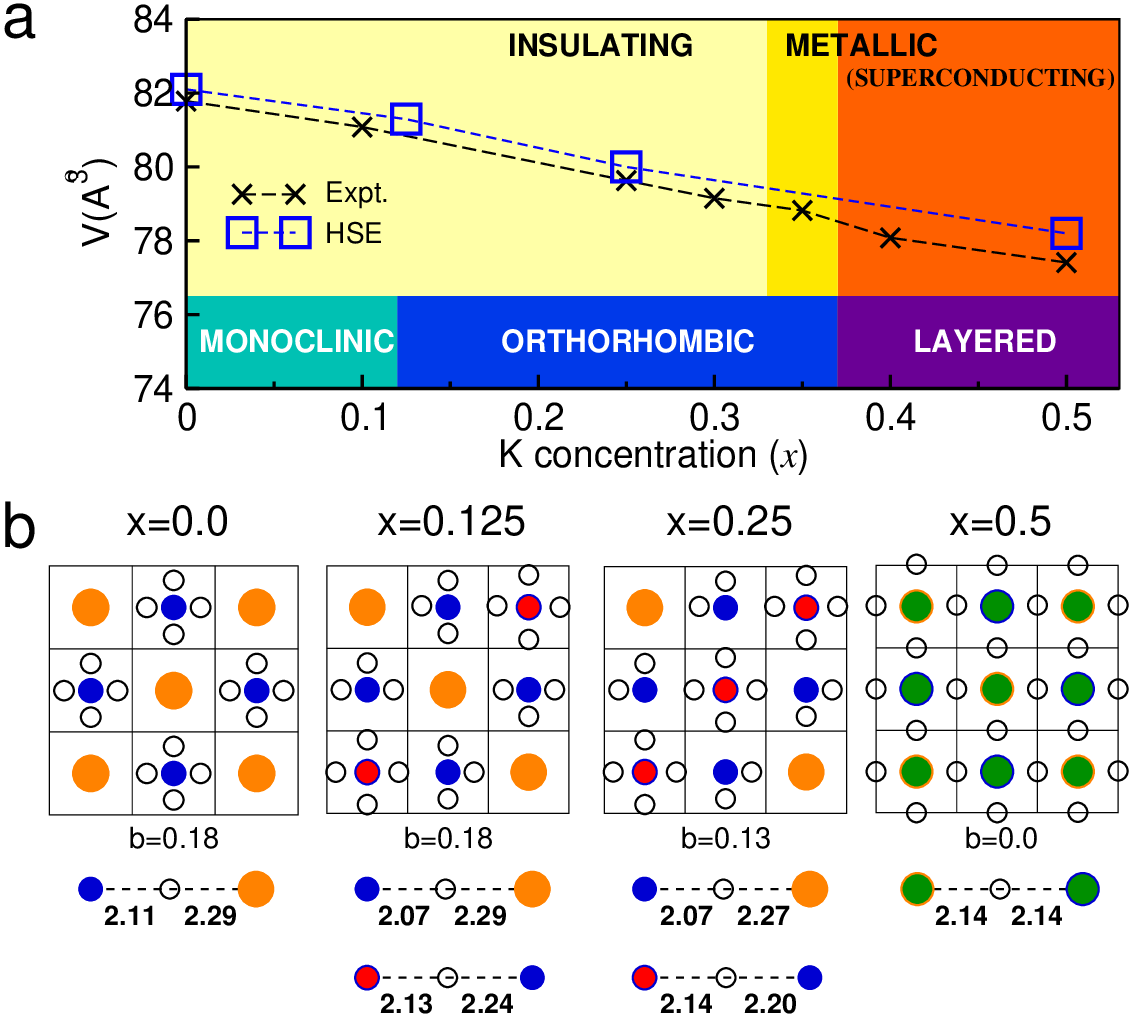}
\caption{
(a) Phase diagram and volume ($V$) of $\rm Ba_{1-x}K_xBiO_3$ upon doping throughout the insulator-metal transition.
The computed volumes are compared to neutron powder diffraction data\cite{pei}.
(b) Schematic structure of the Bi-O plane for the insulating charge ordered state of pure
BaBiO$_3$ ($x=0$), the bipolaronic phases at $x=0.125$ and $x=0.25$ and the metallic regime at $x=0.5$,
showing the local oxygen ($\circ$) breathing environment around $\rm Bi^{3+}$ (\color{Orange} {$\bullet$}\color{black}),
$\rm Bi^{5+}$ (\color{blue} {$\bullet$}\color{black}), bipolaron
$\rm Bi^{3+} \to \rm Bi^{5+}$ (\color{red} {$\bullet$}\color{black}), and $\rm Bi^{4.5+}$
(\color{green} {$\bullet$}\color{black}) sites. 
Bi-O bond lengths between different sites are displayed. 
}
\label{fig:3}
\end{figure}

{\em The non-polaronic state}  was modelled using a supercell comprising 8 pervoskite unit cells [Fig. \ref{fig:1}(b)]. 
For the initial structures--- before relaxing
the structural parameters ---the BiO$_6$ octahedra were tilted in accordance 
to the monoclinic structure of Fig. \ref{fig:1}(a). 
In this supercell, 
we replaced 2 or 4 of the original 8 Ba atoms by K atoms, corresponding to a hole 
doping concentration of $x=0.25$ and $x=0.5$, respectively.

\begin{figure*}
\centering
\includegraphics[clip=,width=0.9\linewidth]{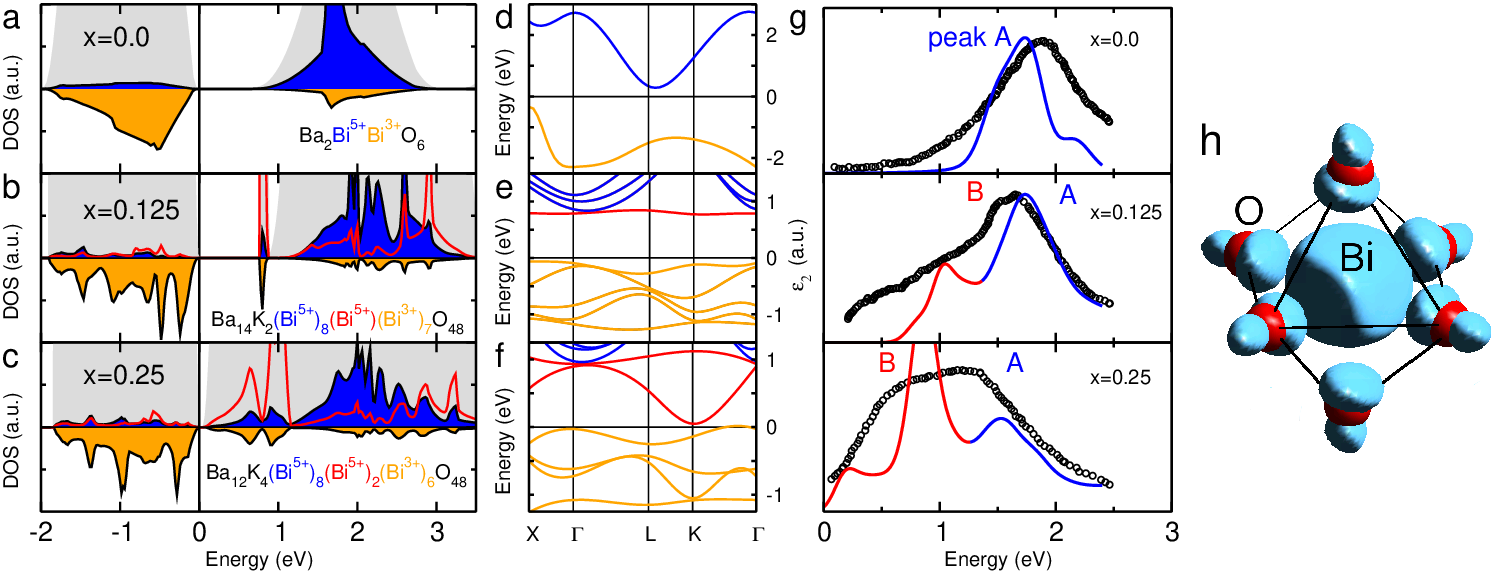}
\caption{
(a-c) Evolution of the DOS and (d-f) corresponding bandstructure
upon K substitution for the polaronic solution.
The gray shadows indicate the total density of states, and red curves indicate the bipolaronic states as created by doping.
(g) Comparison between the theoretical and measured imaginary part of the dielectric function for $x$=0, 0.125 and 0.25.
Peak (A) corresponds to excitations from Bi$^{3+}$ into Bi$^{5+}$ states, peak (B) corresponds to excitations from
Bi$^{3+}$ into bipolaronic states [red curves in panel (b),(c),(e), and (f)].
The experimental curves ($\circ$) are taken from Ref.~\onlinecite{nishio} ($x=0$) and Ref.~\onlinecite{ahmad} ($x=0.21$).
For the calculated curves,  the excitation energies  have been scaled by a factor of 0.8 to approximately account for excitonic effects
otherwise not included in the calculations. 
(h) Charge density corresponding to the bipolaronic band (red line) of panel (e) localised
around the BiO$_6$ octahedron at the converted Bi$^{3+}\rightarrow$Bi$^{5+}$ atom.
}
\label{fig:4}
\end{figure*}

For $x=0.25$, the calculations predict that the two substitutional K atoms prefer to be arranged along
the [001] direction resulting in an orthorhombic supercell after relaxation. 
The predicted lattice parameters agree well with
the experimentally observed orthorhombic structure at $x=0.25$ 
(theory: $a=6.11$~\AA, $b=6.087$~\AA,
$c=8.606$~\AA; experiment\cite{pei}: $a=6.098$~\AA, $b=6.086$~\AA, $c=8.584$~\AA). 
For $x=0.5$, K layers parallel to the (100) planes are preferred.
The final relaxed structure is non-centrosymmetric and tetragonal, and the structural parameters 
again agree very well with the experimental values (theory: $a=b=4.27$,  $c/a=2.01$,
experiment\cite{klinkova}: $a=b{\approx}4.26$, $c/a{\approx}2.0$). 

The density of states shown in Fig. \ref{fig:2}(b) shows that the 
Fermi-level moves into the Bi$^{3+}$ dominated valence band, and therefore 
the system becomes metallic. 
Furthermore, as a result  of the reduced charge 
disproportionation, the Bi$^{3+}$ states move upwards in energy until the
valence and conduction bands overlap  [Fig. \ref{fig:2}(c)].
At $x=0.5$ charge disproportionation is no longer sustained 
\ref{fig:2}(c).
We also simulated this effect by electronic hole doping in the primitive cell (Ba$_2$Bi$_2$O$_6$)
finding quantitatively similar results as for K induced hole doping [Fig. 2(d)],
indicating that chemical effects do not play a major role.

Although the above description accounts well for the structural properties and
the change in volume [see Fig. 3(a)],
the Fermi-level
is located in the valence band for any sizable hole concentration, 
resulting in a strong intra-band metallic behaviour,
in contradiction with experiment\cite{nishio,ahmad}.
On the other hand, polarons
allow the stabilisation of an insulating  groundstate.

{\em The isolated polaron}
was modelled in an fcc like supercell with 16 $\rm BaBiO_3$ unit cells [Fig. \ref{fig:1}(c)] 
replacing two Ba atoms by K atoms ($x=0.125$). 
The BiO$_6$ octahedra were again initially tilted as in the monoclinic structure.
Within this setup the trapping of the two holes on one former Bi$^{3+}$ lattice site
emerges naturally during relaxation.
Remarkably, hole trapping and local lattice relaxations are strongly linked. 
Hence we encounter a truly polaronic behaviour. Since 
a single  Bi$^{3+}$ cation captures two holes, we
will  refer to this Bi lattice site as ``bipolaron'' from now on\cite{ziman,bischofs}. 
The final relaxed model is clearly insulating [Fig. \ref{fig:4}(b)] with an electronic structure that
is  similar to the one of pure BaBiO$_3$. The two holes are confined in 
one single unoccupied band below the bottom of the conduction band [Fig. \ref{fig:4} (e)]. 
The hole trapping is accompanied by a fairly large local relaxation and
since converted sites have only Bi$^{5+}$ neighbours,
they  are structurally frustrated and do not order within the same local breathing environment as 
the non-frustrated original Bi$^{5+}$ atoms, 
as depicted in detail in Fig.\ref{fig:3}(b). 

The spatial charge distribution of the narrow bipolaronic band  is shown in Fig. \ref{fig:4}(h).
Only about 15 \% of the charge is localised on the Bi site, 
the rest is partially located on the surrounding O atoms and a significant amount of charge is also
dispersed throughout the cell (not shown). This distinguishes $sp$ systems from
$d$ or $f$ dominated materials. Fig. \ref{fig:4}(h) shows that the band is made up by
an antibonding linear combination between the Bi $s$ orbital and the surrounding O $p$ orbitals (nodal planes along Bi-O bonds).
Hence the lattice distortion accompanying the bipolaron is driven by moving 
the antibonding Bi-O state above the Fermi-level. The bonding state is occupied and
found in the O subband. This  results in the observed contraction of the O-Bi bond length 
for the converted Bi site (2.13 \AA)
with respect to the non-converted Bi$^{3+}$ sites (2.29 \AA). 
Attempts to stabilise such a solution using more approximate techniques such as 
a one-centre LDA+U treatment failed for this material, since the Bi $s$ orbitals are too delocalised.

{\em Interaction between polarons:}
stabilising the polaronic solution at a larger hole concentration ($x=0.25$) turned out
to be rather difficult. To minimise bipolaron-bipolaron interactions, one might naively expect the bipolarons to
arrange in a sc $2\times 2 \times 2$ super-lattice. 
But relaxation from such a starting structure always turned into a metallic state [see Fig. \ref{fig:1}(b)].
In the fcc like supercell with 16 $\rm BaBiO_3$ unit cells, however, 
a brief and completely unbiased molecular dynamics run, followed by relaxation yields an
insulating state. A string of Bi$^{3+}$ atoms along the face diagonal [101]
converts into Bi$^{5+}$ atoms, each trapping two holes [Fig. \ref{fig:1}(c)].
Similar bipolaronic alignments along
all three face diagonals are found to be energetically almost degenerate, 
all leading towards a similar insulating state as displayed in Figs. \ref{fig:4}(c) and (f). 
Compared to the metallic state the insulating solution is about 150 meV/polaron lower in energy.
The sc arrangement of bipolarons [Fig. \ref{fig:1}(b)] was consistently metallic
for symmetry reasons (lack of some Bragg reflections for this arrangement) \cite{ordering}.
As a result of the small distance between the 
bipolarons, we find a rather strong electronic interaction between them.         
The bipolaronic subband [Fig. \ref{fig:4}(c)] is split into a lower broad  band  and a relatively
narrow second subband [Fig. \ref{fig:4}(f)].
The observed band structure is consistent with weakly interacting
$s$ like impurity states, with the lower (upper) subband corresponding to bonding (anti-bonding) linear combinations 
of bipolaronic states.
Although the interaction between bipolarons will partially
depend on  the long range order, the strong interaction beyond $x=0.125$
might well be the origin of the experimentally observed monoclinic to orthorhombic
phase transition at $x\approx0.12$. The strong interaction ultimately
also causes an overlap between the bipolaronic band and the valence band resulting
in the metallic non-disproportionated state observed in our calculations at $x=0.5$ [Fig. \ref{fig:2}(c)].
A proper description of the superconducting state emerging at these high doping concentrations
will require the inclusion of electron-phonon coupling. For the insulating state, however,
we expect only minor modifications due to electron-phonon coupling.

The comparison between the experimental 
and the calculated imaginary part of the dielectric function ($\epsilon_2$) shown in Fig. \ref{fig:4}(g)
confirms that our model captures all essential features.
In pure $\rm BaBiO_3$ a single adsorption peak (A) is visible corresponding to the charge-ordered excitation between
the Bi$^{3+}$ and Bi$^{5+}$ sub-bands. At $x=0.125$, a second peak (B) emerges related to
the excitation from the valence Bi$^{3+}$ band
into the  bipolaronic band. It should be noted that disorder 
and finite temperature effects (electron-phonon interaction)
will broaden the theoretical peaks thus improving the 
agreement with experiment.
At $x=0.25$, the bipolaronic peak (B)
increases significantly in intensity and shifts towards lower energy,
whereas peak A loses intensity.
Considering that our calculations neglect disorder and approximate excitonic effects, our
results compare well with experiment and qualitatively explain the 
emergence of peak B in the experimental spectra, its low intensity at low doping concentrations, 
the downshift and increase in intensity at larger doping concentrations
(bipolaron-bipolaron interaction), 
as well as the bleaching of the A mode.

In summary, we have shown that $sp$ like electronic holes can trap 
at otherwise perfect lattice sites, if the lattice is sufficiently flexible
and capable of screening the local hole.  
One important observation of the present study is that $sp$ mediated polarons are not particularly
well localised (only 15\% of the charge localised at the Bi polaron site, see Fig. \ref{fig:4}(h)).
Resultantly, at higher concentrations we find bonding and anti-bonding 
bipolaronic states and strong ordering tendencies.  For $sp$ systems the opening of
a gap is more akin to a "long-range" Peierls distortion than to an essentially ``local'' Jahn-Teller effect.
Similar features can be expected
for many multivalent cations. 
From a computational point, 
our study demonstrates that hybrid functionals describe multivalent cations 
as well as the resultant polaronic 
lattice distortions outstandingly well, 
and represent a major and ``not just quantitative'' step forward in the modelling of oxides. 
We believe that the successful description of the insulating oxide Ba$_{1-x}$K$_x$BiO$_3$
will spur similar research in a wide class of complex physical phenomena such as
trapping of electrons and holes in high-k dielectric materials or organic semiconductors, 
or confinement at the interface between metallic and insulating oxides.


\begin{thebibliography}{99}
\bibitem{ordering}
To clarify why the sc arrangement of bipolarons is unstable, we inspected
the bandstructure of the metallic cubic cell  [Fig. \ref{fig:1}(b)], and found
three valence bands with maxima at $\Gamma$ (empty) and strong and similar 1.5 eV downward
dispersion along $\overline{\Gamma \mathrm X}$ and  $\overline{\Gamma \mathrm L}$ (states at X and L are occupied).
Although this supercell is
orthorhombic,  the original cubic symmetry of the perovskite unit cell prevails
to a large extend in the bandstructure. A semiconducting state can be expected only
if states at $\Gamma$ couple with those at L and/or X in the 8-fold supercell.
At $x=0.125$ an fcc like arrangement fullfills this criteria, whereas
at $x=0.25$, the  second  bipolaron must be placed such that the Bragg
reflections prevail along space diagonals. This can be realised only
by ordering all bipolarons in stripes along
the face diagonals, as indicated by small red circles in  Fig. \ref{fig:1}, whereas
bipolaronic conversion of the body centred atom, circled green in Fig. \ref{fig:1}(c),
destroys the Bragg reflection along the space diagonal.
\end{thebibliography}
\end{document}